\documentstyle[12pt, a4wide]{article}

\def\bg{\begin{equation}}
\def\ed{\end{equation}}
\def\bq{\begin{eqnarray}}
\def\eq{\end{eqnarray}}
\def\eps{\varepsilon}
\def\dhdy{{\partial H \over \partial y}}
\def\dhdx{{\partial H \over \partial x}}

\def\Hb{\bar{H}}
\def\tb{\bar{t}}
\def\deH{\Delta H}
\def\oe{{\cal O} (\eps)}
\def\oes{{\cal O} (\eps^2)}

\def\tP{\tilde{P}}

\def\DHg{DH \cdot g}
\def\cosT{\cos{({2\pi \over T} t_0)}}

\def\spr{separatrix \/}
\def\stch{stochastic \/}

\newcommand{\csch}[1]{\mbox{csch} \left( #1 \right)}

\begin{document}
\newfont{\FieldFont}{msbm10 scaled\magstep1}
\newcommand{\FZ}{\mbox{\FieldFont Z}}  
\newcommand{\FR}{\mbox{\FieldFont R}}  
\baselineskip=24pt

\title{Chaotic Transport in Planar Periodic Vortical Flows}
\author{Taehoon Ahn \\ and \\  Seunghwan Kim \\
Departments of Mathematics and Physics, \\
Basic Science Research Institute \\ 
POSTECH \\
 P.O.Box 125, Pohang, Korea}
\date {}
\maketitle

\begin{abstract}

        We have studied \vspace*{0.3cm}a chaotic transport in a two-dimensional  periodic 
vortical flow under a \vspace*{0.3cm} time-dependent perturbation with period $T$, where the 
global diffusion occurs along \vspace*{0.3cm} the stochastic web. By using the Melnikov
method we construct the separatrix map describing the approximate \vspace*{0.3cm} dynamics near
the saddle separatrices. Focusing on the small $T$, the width of the \vspace*{0.3cm} stochastic 
layer is calculated analytically
by using the \vspace*{0.3cm} residue criterion and the diffusion constant 
by using the random phase assumption and \vspace*{0.3cm} the theory of correlated
random walks. The analytical results are in good agreements with the results 
of \vspace*{0.3cm} two different types of numerical simulations by integrations of
the \vspace*{0.3cm} Hamilton's equation of motion and by iterations of the separatrix map,
which establishes the validity of the use of the separatrix map.

\end{abstract}


\section{Introduction}
Recently the global transport and mixing in the problems of planar fluids have
been studied intensively by using the framework of the dynamical
system theory, where interesting physical quantities  like mass or heat can be
described as moving with the fluid particle to some degrees of approximation 
\cite{Wss, Och, Cmss}.

When the flow is laminar,
the velocity vector field $v(x,y,t)$ of a two dimensional incompressible
inviscid fluid flow can be determined from a stream function $\Psi(x,y,t)$
where \bg
        v(x,y,t) = \Bigl({\partial \Psi \over \partial y},
                        - {\partial \Psi \over \partial x}\Bigr).
        \ed
The equations for an infinitesimal fluid element
        (called a {\em fluid particle}) become
        \bg
        \dot{x} = {\partial \Psi \over \partial y} (x,y,t) \>\>\>\>\>
        \dot{y} = -{\partial \Psi \over \partial x} (x,y,t).
        \ed
This is simply a Hamiltonian dynamical system where the stream function plays
the role of the Hamiltonian. It is now well established that
for the time-dependent stream function
the particle trajectories can display chaotic dynamics, even though the
Eulerian flow is laminar \cite{Ott}. Such a flow is called to exhibit the
{\em chaotic advection}. In general, the transition from the time-independent
to the time-dependent flow corresponds to the change from an integrable to
a near-integrable Hamiltonian system, which may turn globally connected
separatrices into stochastic layer, so that the fluid particles within these 
stochastic layer can be transported globally. 

In this paper we consider the planar periodic vortical flow in which there
are four hyperbolic saddle points connected by heteroclinic orbits
\cite{Brtzz}. This is an interesting model for certain convection problems as 
well as the axisymmetric Taylor vortex flow and the Rossby waves of geophysical
fluid dynamic \cite{Pdl}. 
With a time-periodic perturbation the heteroclinic saddle 
connections (or separatrices in terms of the Hamiltonian system) break down and
form a globally connected stochastic layer (called the {\em stochastic web}).
Fluid particles diffuse chaotically along the stochastic layer, which can be regarded as
a stochastic process in a coarse-grained scale. This stochastic transport
occurs on the global scale along
the 2-dimensional square-like lattice of the stochastic web.

The {\em separatrix map} is an approximate map describing the dynamics of the energy
and the phase near the separatrices of the periodic Hamiltonian system, 
first introduced by Chirikov \cite{Chrv}, which allows an
efficient and systematic study of dynamics near the stochastic web. Escande was 
first used the separatrix map for the transport problem \cite{Esc} and Weiss
and Knobloch used the separatrix map to study numerically the anomalous 
diffusion of the fluid particle along modulated traveling waves \cite{Wss}.
In our paper, we follow Weiss and Knobloch to
construct the separatrix map for the periodic vortical flow and use it
to estimate the width of the stochastic web and the 
diffusion constant analytically, extending the analysis of Lichtenberg and Wood
on the Hamiltonian system of charged particles in a magnetic field \cite{Lch}.

In spite of the wide usage of the separatrix map
for studying the transport problems \cite{Esc, Lch, Afn}, however,  the validity
of the separatrix map analysis has not been firmly established yet. 
One of the aims of this paper is testing the validity of 
the separatrix map as a general method for studying transport in
the Hamiltonian system as well as studying the transport in planar fluids.
We will show that if 
the period of the perturbation $T$ is small, the \stch layer is sufficiently 
wider than the splittings of stable and unstable manifolds and the \spr map 
can describes the structure of the \stch
layer apart from the \spr very well. Moreover, as $T$ decreases the advantage 
of using the \spr map increases since one iteration of the \spr map 
corresponds to a
large number of iterations of the time-$T$ Poincar\'{e} map, so that it becomes 
much easier to study asymptotic dynamics with the former than the latter.
The problem may arise when the orbit approaches the separatrix 
since the error of the separatrix map grows indefinitely \cite{Rkd}.  
We will show that this can 
be overcome by assuming the phase as a random variable for dynamics near 
the separatrix.  

In our problem, the separatrix map provides
a first order approximation to true dynamics, with which we were able to 
calculate the analytic form for the width of the stochastic layer, the mean 
square distance of fluid particles, and the global diffusion constant. The 
{\em random phase assumption} on the separatrix map suggests that the \stch
process governing the global dynamics from one cell to another can be described 
by a two-dimensional {\em correlated random walk} \cite{HsKhr}, from which 
a global diffusion constant can be computed. 

First, we compute the analytical form of the \spr map
by introducing a set of Poincar\'{e} surface of sections. We find that 
the Greene's residue criterion \cite{Grn} gives a quite accurate estimate for 
the width of the \stch layer. We introduce a random model describing dynamics
of the energy by assuming the phase as a random variable, which is 
different from the model of Lichtenberg and Wood, where both the energy and phase
are random variables. This assumption leads to the correlated random walks in
a coarse grained level, with which we can obtain an analytic estimate of the
diffusion constant. We make a correction to the diffusion due to elliptic 
islands in the \stch layer. 

        In section 2, we introduce the model Hamiltonian describing the flow 
with the vortex lattice with saddle-connections under time-periodic 
perturbations. We show the formation of the globally connected stochastic web by computing 
the Melnikov function and applying the Smale-Birkhoff theorem. We also 
construct the separatrix map and provide an error analysis on the separatrix 
map. In section 3, we describe the structure of the global stochastic layer and 
calculate the analytical form for the width of the stochastic layer by using
the residue criterion on the \spr map. In section 4, based on the random phase
assumption, the diffusion constant 
is computed by applying the theory of correlated random walks. We compare 
analytical results with those from simulations. Finally, we end with concluding remarks.

\section{The Melnikov method and the separatrix map}
\subsection{A model}

As a model for a two-dimensional time-dependent flow we consider the following 
stream function \cite{Brtzz}:
	\bg
	\Psi (x,y,t) =  H(x,y) + \eps H_1(x,y,t)
	\ed
where
	\bq
	H(x,y) &  = & {1 \over 2\pi}\sin{(2\pi x)} \cos{(2\pi y)}  \cr
	H_1(x,y,t) &  =  & {1 \over 2\pi}\cos{(\omega t)} 
		[\sin{(2\pi y)} + \cos{(2\pi x)}]. 
	\eq
This stream function corresponds to the near-integrable Hamiltonian system with
time dependent perturbation of period $T = {2\pi \over \omega}$ and the 
equations of motion for fluid particles are given by
	\bg
	\pmatrix{ \dot{x} \cr \dot{y} \cr} =
	\pmatrix{-\sin{(2\pi x)}\sin{(2\pi y)}+\eps\cos{(2\pi y)}\cos{(\omega t)}\cr
	-\cos{(2\pi x)}\cos{(2\pi y)}+\eps\sin{(2\pi x)}\cos{(\omega t)}\cr}
	\ed
or, in a vector form,
	\bg
	\dot{q} = JDH_0(q) + \eps g(x,y,t), \nonumber
        \ed
where $$
	 q = (x, y), \>\>
	 J \equiv \pmatrix{0& 1\cr  -1&  0\cr}, \>\> 
	 DH \equiv \pmatrix{ {\dhdx} \cr 
		 {\dhdy} \cr } \hbox{,  and} $$
	$$ g(x,y,t) = JDH_{1}(x,y,t). $$
This equation describes a two-dimensional periodic vortical flow and its time 
dependent perturbation.
This flow is obviously doubly periodic, yielding a flow on the torus. Viewed
as a flow on the torus, the unperturbed	system is an integrable Hamiltonian
system with heteroclinic orbits connecting four saddle points. On the torus,
there exist four vorticies whose boundaries are the heteroclinic 
orbits. The values of $H$ are zero on the heteroclinic orbits and have a 
definite sign for each vortex. A heteroclinic orbit is both a stable manifold 
of one saddle point and an unstable manifold of another saddle point. The phase
portrait of the unperturbed system is shown in Figure 1. Under the periodic 
perturbation the heteroclinic connections break down and the stable and 
unstable manifolds between two adjacent saddle points in general do not 
coincide with each other and intersect transversally. 

\subsection{The Melnikov method}
	For a system with a saddle connection under a time-dependent 
perturbation Melnikov devised a method for finding the transverse intersection 
of stable and unstable manifolds \cite{Mlnkv}. Let $q_s(t)$ be the heteroclinic 
orbit of the unperturbed system. If we parametrize the heteroclinic
connection by $t_0 \in \FR$, the distance between the stable and unstable manifolds
along the direction normal to the unperturbed heteroclinic orbit at $q_s(-t_0)$,
denoted by $d(t_0, \eps)$, is given by
	\bg
	d(t_0, \eps) = \eps {M(t_0) \over \parallel DH(q_s(-t_0))\parallel}, 
	\ed
where $\parallel\cdot\parallel$ is the Euclidean norm in $\FR^2$ and $M(t_0)$ 
is the Melnikov function 
	\bg
	 M(t_0) = \int^{\infty}_{-\infty} \DHg(q_{s}(t),t+t_{0})\,dt.
        \ed
The Melnikov theory says that if the Melnikov function
has a simple zero, then the stable and unstable manifolds intersect transversely
at $q_s(-t_0) + \oe$ \cite{Gck}.

	The Melnikov function for our Hamiltonian vector field can be 
computed explicitly. Along all unperturbed horizontal connections we have
	\bg
	\sin{(2\pi x)} = \pm {2\exp{(-2\pi t)} \over 1 + \exp{(-4\pi t)}}
	\ed
and along all vertical connections
	\bg
	\cos{(2\pi y)} = \pm {2\exp{(-2\pi t)} \over 1 + \exp{(-4\pi t)}}. 
	\ed
Then the Melnikov function in Equation (8) becomes  
	\bq
	M^{\pm}(t_0) & = & \pm\int^{\infty}_{-\infty} {\exp{(-4\pi t)} 
		\over (1 + \exp{(-4\pi t)})^2}\cos{\omega (t+t_0)}\,dt \cr
	& = & \pm {\omega \over 4\pi}{1 \over \sinh{({\omega\over 4})}}
		\cos{(\omega t)} 
	\eq
or
	\bg
	M^{\pm}(t_0) = \pm M_0(T) \cosT, 
	\ed
where $$
	M_0(T) = {1 \over 2T} \csch{\pi\over 2T}. 
	$$
Hence we see that the Melnikov function is periodic with period $T$ and
has two simple zeros per period and, 
therefore, by the Melnikov theory the stable and the unstable manifolds 
intersect transversely. If the system has a transversal intersection of the 
stable and the unstable manifolds, the Smale-Birkhoff homoclinic 
theorem provides an existence of chaotic orbits \cite{Smale, Bkf}. This theorem 
is naturally extended to the heteroclinic case \cite{Brtzz}, which 
proves that our system has a chaotic orbit under a periodic perturbation 
$H_1(q,t)$. In fact, the chaotic orbits form a \stch layer along the 
separatrices and the fluid particle can be globally diffused along the connected
stochastic layer. We note that the T-dependence of $M_0(T)$ is of the form, 
$\exp{(-{1\over T})}$, as proved by Holmes et al. \cite{Hml}. Therefore, 
the separatrix splittings are exponentially small in the small T-limit, 
yielding the exponentially thin stochastic web. 

\subsection{The separatrix map}
The dynamics near the separatrices of the periodic Hamiltonian system
can be approximated by the separatrix map. 
	In order to construct the separatrix map, we choose two saddle points 
$q_0 = (0,{1\over 4})$ and $q_1 = (0,{3\over 4})$ connected by the heteroclinic orbit
	\bg
	q_s(t) = \Biggl(0, {1\over 2\pi} 
	\arccos{\Bigl(-{2\exp{(-2\pi t)} \over 1 +\exp{(-4\pi t)} } \Bigr) }\Biggr).
	\ed
We also choose three different surfaces of section:
	\bq
        \Sigma^{t} & = & \Bigl\{ (x,y) \mid \> \mid x\mid < \delta,y = {1 \over 2}\Bigr\} \cr
        \Sigma^{H}_0 & = & \Bigl\{ (x,y) \mid \> \mid x\mid <\delta, y={1\over 4}+\mid x\mid\Bigr\} \cr
        \Sigma^{H}_1 & = & \Bigl\{ (x,y) \mid \> \mid x\mid <\delta, y={3\over 4}-\mid x\mid\Bigr\}, 
        \eq
where $\delta$ is a small positive value. Suppose that these sections are 
chosen for each separatrix. Suppose also that $q(t)$ is a perturbed orbit 
such that $q(s_n)\in \Sigma^{H}_0, \>\> q(t_n)\in \Sigma^{t}, \>\> 
		q(s_{n+1})\in \Sigma^{H}_1 $
and $q(t)$ crosses  $\Sigma^{t}$ of the next separatrix at time $t_{n+1}$, 
as shown in Figure 2. The
change of the energy $H$ from $q(s_n)$ to $q(s_{n+1})$ along $q(t)$ is given by
	\bg
	\deH = \int^{s_{n+1}}_{s_n} {dH \over dt}\, dt
		= \eps \int^{s_{n+1}}_{s_n}\DHg (q(t),t)\, dt.
	\ed	 
From the relation ${dx\over dt} = \dhdy, \>\>\> t_{n+1}$ is given by
	\bg
	t_{n+1} = t_n + \int^{q(t_{n+1})}_{q(t_n)}\dhdy (q(t),t)^{-1}\, dx.
	\ed
We call the map $S_e$ : $(H,t_n) \rightarrow (H+\deH,t_{n+1})$ the {\em exact 
separatrix map}, which describes the exact dynamics of the energy $H$ and the
time (or phase) $t$. Here $(H,t)$ plays the role of a new coordinate system in the 
planar fluid space near the separatrix.
Now we approximate $q(t)$ by the heteroclinic orbit $q_s(t)$ and $t_{n+1} -t_n$
by the period of an unperturbed solution  $q_0(t)$ with $H_0 = H + \deH$. Then
$\deH \simeq \eps M^{\pm}(t_n)$, where $M^{\pm}(t_n)$ is the Melnikov 
function in Equation (12). Then an (approximate) separatrix map is given by
        \bg
        \pmatrix{H_{n+1} \cr t_{n+1}\cr} = S^{\pm}(H_{n}, t_{n}) =
                \pmatrix{H_n + \eps M^{\pm}(t_n)\cr
                        t_n + {1\over 4}T(H_n + \eps M^{\pm}(t_n))\cr},
        \ed
where $T(h)$ is a period of an unperturbed orbit with $H = h$. It can be 
shown that
	\bg
	S_e(H_n, t_n) = S^{\pm}(H_n,t_n) + \pmatrix {\oes \cr \oe}
	\ed
for $\mid H_n \mid = \oe$, i.e., $S^{\pm}$ is valid for small $\eps$, that is, 
near the 
separatrix \cite{Rkd}. We note that if $T$ is small, then the change of $H$ 
is exponentially small, so that the validity of $S^{\pm}$ does not appear to be 
justified. However, the numerical observation shows that $\oes$ error term 
is also exponentially small, justifying the use of $S^{\pm}$. However, this 
issue is a mathematically difficult problem and is not fully understood yet 
\cite{Hml, Fnt, Mrsd}. 

	Near the separatrices a straightforward computation  for our system in
Equation (3) yields
	\bg
	T(H) = {2 \over \pi}K(1-4\pi^2H^2) \simeq 
	{2 \over \pi}\ln{2 \over \pi\mid H\mid} 
		\mbox{ \hspace{0.5cm} for \/} \mid H\mid \ll 1,
	\ed
where $K(\cdot)$ is the complete elliptic integral of the first kind. Therefore, 
we get the separatrix map for our system 
	\bg
	\pmatrix{H_{n+1}\cr t_{n+1} \cr} = S^{\pm}(H_n,t_n) = \pmatrix{
	H_n \pm \eps M_0(T)\cos{({2\pi\over T} t_n)} \cr
		t_n + {1\over 2\pi} \ln{2 \over \pi \mid H_{n+1}\mid} \cr},
	\ed
which is used throughout the  rest of the paper to explore the chaotic 
transport.

\section{The structure and the width of the stochastic layer}
\subsection{The \spr map versus the twist map}

In order to study the dynamics in the stochastic layer, it is important to
understand the structure of the stochastic layer first. The stochastic layer 
is composed of not only chaotic orbits but also other complex structures such 
as elliptic islands and cantori which exist away from the 
separatrices and become relevant to transport when $T$ is small. These 
structures can be described very well by the separatrix map. From Equation (20) 
we see that the value of $H$ changes roughly by the amount
$\eps M_0(T)$ as a fluid particles moves along a separatrix. When $T$ is small, 
we will show later that the width of the stochastic layer is significantly 
larger than $\eps M_0(T)$. Therefore the values of $H$ have a definite sign for a long 
time even for chaotic orbits, which corresponds to oscillating  dynamics
confined in one cell. In order to study oscillations in a single cell, it is 
convenient to introduce the {\em composite separatrix map} 
$S = S^+ \circ S^- \circ S^- \circ S^+$ because one 
iteration of $S$ corresponds to one oscillation confined in a cell {\em if} the 
sign of $H$ does not change. Since $M_0(t)$ is $T$-periodic, the map is 
defined on the cylinder with a form similar to a {\em twist map}, such as the 
{\em standard map}:
	\bq
	y^\prime & = & y + {k \over 2\pi} \sin{2\pi x} \cr
	x^\prime & = & x + y^\prime. 
	\eq 
Though the map $S$ is similar to the twist map, it is different in the 
sense that it has a logarithmic 
singularity at $H=0$ and its twist conditions depend on 
the sign of $H$. Away from the separatrix but still within the stochastic 
layer, however, $S$ can be treated as a twist map, so that the theory of 
twist maps can be applied to our problem. From Equations (20) and (21) we know that the dynamics
of $S$ for small $|H|$ is similar to that of the standard map for large 
$k$ because the variations in $t$ and $x$ increase as $|H|$ 
decreases or $k$ increases. The map $S$ bears strong resemblances to 
the standard map in this aspect.

	It is well known that the twist map has complex 
dynamical structures with periodic orbits, 
quasi-periodic orbits forming invariant circles, chaotic 
orbits and {\em cantori} 
\cite{Mss}. The cantori 
are the invariant Cantor sets in which the motion is quasi-periodic and they 
typically exist in the stochastic layer. A general proof of the existence of 
cantori for the twist map has been given by Aubry et al. \cite{Abr}, Mather 
\cite{Mth} and Katok \cite{Ktk} and it guarantees that cantori exist in the
region for which $S$ can be regarded as a twist map. 
In the cantori there exist an infinite number of gaps whose total length is finite,
so that most of them are very small. Even when there are large gaps, fluid 
particles can take a long time to get through. So the cantori are partial 
barriers for the transport of the fluid particle.

\subsection{The rotational invariant circle and the residue criterion}
 If an invariant circle is a closed loop that encircles the cylinder we call 
this the {\em rotational invariant circle \/} (RIC). The boundaries of the 
stochastic layer are the outer most RIC of $S$, which determines the width of the
stochastic layer. We will see later that the calculation of the width is 
important in calculating the diffusion constant of the global transport along 
stochastic layer. Birkhoff showed that RIC of the twist map is a Lipschitz 
graph $\{Y(t),t\}$ of some continuous function $Y$ \cite{Bir}. Using the fact
that RIC is a Lipschitz graph, Mather showed that for the standard map, there
is no RIC if $|k| > 4/3$. An application of his method to our problem gives 
the following rigorous lower bound for 
the half width $W_h$ 
	\bg
	W_h \geq {3\over 8}{\eps \over T^2} \csch{\pi \over 2T}.
	\ed
But in order to calculate the diffusion constant the above lower bound is not 
sufficient. Therefore, in order to estimate the width more accurately 
we use the residue criterion, introduced by Greene \cite{Grn}.
Given a  periodic orbit of period $q$, its residue $R$ is defined by \bg
	R = {1\over 4} ( 2 - Tr[DS^q]),
	\ed
where $DS^q$ is the Jacobian matrix of the $q$-th iteration of the map $S$.
For the twist map it is known that for each rational $p/q$ there exists at
least one periodic orbit of rotation number $p/q$ with non-negative 
residue $R_{p/q}$. The residue criterion for the existence and nonexistence of 
\vspace{10pt}
RIC is given as follows: \\
{\bf Criterion} {\em Given two neighboring rationals $p/ q$, $p^{\prime} / 
	q^{\prime}$, i.e., $pq^{\prime} - qp^{\prime} = \pm 1$, there exists no
	rotational invariant circles of rotation number between $p/ q$ and
	$p^{\prime} / q^{\prime}$ if the residue $R_{p/q}$ and 
	$R_{p^{\prime}/q^{\prime}}$ are significantly greater than 0.25. If the
	residues are significantly smaller then such circles do exist.}
\vspace{10pt}

\subsection{The estimate of the width of the \stch layer}
The composite separatrix map $S$ has infinitely many fixed points of rotation 
number $2m/1$ near the separatrix $(\Hb, \tb)$, satisfying 
	\bg
	M^{\pm}(\tb) = 0 \mbox{\hspace{1cm} and \hspace {1cm}} T(\Hb) = 2mT,
	\ed
where $m$ is a large positive integer, or, equivalently
	\bg
	\tb = {T\over 4} \mbox{\hspace{0.2cm}or\hspace{0.2cm}} {3\over 4}T 
		\mbox{\hspace{1cm} and \hspace {1cm}}
		\Hb = {2\over \pi} \exp{\left( -{m\pi T \over 2} \right)}.
	\ed
The map $S$ also has fixed points of rotation number $(2m+1)/1$ near 
$(\Hb, \tb)$, where
	$$
	M^+(\tb) + M^-(\tb+{mT\over 4}) + M^-(\tb+{mT\over 2}) 
			+ M^+(\tb+{3mT\over 4}) = 0
	$$
and
	\bg
        T(\Hb) = (2m+1)T.
        \ed
These two classes of the fixed points of rotation number $m/1$ correspond to the zeros of 
$M^{m/1} (t)$, the subharmonic Melnikov function \cite{Wig}, since 
	\bg
	M^{m/1} (t) \simeq M^+(t)+M^-(t+{mT\over 4})
	+M^-(t+{mT\over 2}) + M^+(t+{3mT\over 4}), 
	\ed
for sufficiently large $m$.
By using the subharmonic Melnikov theory it can be proved from the implicit
function theorem that a simple zero of $M^{m/1} (t)$ corresponds to a periodic
orbit with period $mT$. Therefore the fixed points of S correspond to periodic
orbits. For the standard map if $k$ is sufficiently large all periodic 
orbits and quasi-periodic orbits of rotation number $p/q$ are 
hyperbolic \cite{Grff}. Similarly, for the 
map $S$, the fixed points of rotation number $m/1$ for sufficiently large $m$ 
are hyperbolic, so that their non-negative residues are greater than $1$. If 
$m$ is not sufficiently large or fixed points are sufficiently far away from 
the separatrix, their non-negative residues are less than 1, that is, they 
are elliptic, so that such fixed points correspond to resonance bands. In 
Figure 3 we plot the phase portrait of $S$ and the phase portrait of the time-T 
Poincar\'{e} map near a saddle point, whose island structures show the
correspondence between the 
resonance bands of the time-$T$ Poincar\'{e} map and the fixed points of $S$. 
For the fixed points of rotation number $2m/1$, we can calculate the residue 
exactly 
	\bg
	R_{2m/ 1} = {1\over 4}(8r - r^2),
	\ed
where \bg
	r = {\left({\eps\csch{\pi\over 2T} \over 2T^2H_{2m/ 1}}\right)}^2
		\mbox{\hspace{0.5cm} and \hspace{0.5cm}} 
			\mid H_{2m/ 1} \mid = {2\over\pi}e^{-\pi mT}.
	\ed
From the condition $R_{2m/1} > 1$, the resonance bands of rotation number 
$2m/1$ should be found in the region of  \bg
	\mid H\mid \>\> > \>\> {\eps\csch{\pi\over 2T}\over {2(\sqrt{3}-1)T^2}}.
	\ed

	From the residue criterion we know that the boundary circle is located
between two elliptic periodic orbits corresponding to neighboring rationals 
with non-negative residue near 0.25. Since in our case the boundary circle 
determining the stochastic layer is located between two elliptic islands of 
rotation number ${2m/ 1},\>$ and ${(2m+2)/ 1}$, which are very close, we can get
an estimate of $W_h$ by applying the residue criterion on $R_{2m/ 1}$ and 
$R_{2m+2/1}$ though ${2m/ 1}$ and ${(2m+2)/ 1}$ are not neighboring rationals. 
The condition $R_{2m/1} \geq {1\over4}$ gives 
the half width of the \stch layer $W_h$
	\bg
	\mid H \mid \geq {\eps \over \sqrt{10} - \sqrt{6}}
			\frac{1}{T^2}\csch{\pi \over 2T} = W_h.
	\ed
We note that the ratio between $W_h$ and $\eps M_0(T)$ is given by  
	\bg
	{W_h \over \eps M_0(T)} = \frac{2}{(\sqrt{10} - \sqrt{6}) T},
	\ed
which indicates that $W_h$ is significantly larger than $\eps M_0(T)$ 
when $T$ is small.
Figure 4 shows the comparison between the above analytical estimate in Equation
(31) and the results of numerical simulations by integrating Equation (5) 
directly and by the iteration of the separatrix map, which shows excellent 
agreements between various estimates over a range of $T$. 

\section{The random phase model and the global diffusion constant}
\subsection{The random phase assumption}

We now consider the problem of the calculation of the global diffusion constant
in the stochastic layer. Typically, a fluid particle in the stochastic layer
displays chaotic oscillations in one cell for some periods before crossing the 
separatrix and displaying chaotic oscillations in a neighboring cell. Therefore
it is natural to consider the global motion in the stochastic layer as the 
random walks over the cells. Lichtenberg and Wood \cite{Lch} 
used this idea to calculate 
the diffusion constant for charged particles in a periodic magnetic field along
the stochastic web with the assumption that dynamics in the stochastic layer 
are completely random, that is, the phase space within the stochastic 
layer is equally populated on each separatrix mapping step. But this assumption
is not adequate for the case of small T because the changes of $H$ per one 
mapping step is limited by $\eps M_0(T)$ which is significantly smaller than
the width of the stochastic layer. As an alternative we consider only $t$ as a
random variable, which is called the {\em random phase assumption}. From 
Equation (20) we see that if $|H| \rightarrow 0$, then $t_{n+1} - t_n 
\rightarrow \infty$, so that this assumption is reasonable near the separatrix. 

	With the random phase assumption we have a random model describing the 
dynamics of $H$ as follows:  
	\bg
	H_{n+1} = H_n + \eps M_0(T) \cos{\xi}, \mbox{\hspace{1cm}} 
		\mid H \mid < W_h
	\ed
where $\xi$ is a random variable in $[0, 2\pi]$. 

\subsection{The diffusion and correlated random walks}
When the sign of $H$ changes 
the fluid particle crosses the separatrix. Since the crossing is regarded as one
random walk step, it is important to know the statistics of the 
separatrix crossing. The random model gives a set of probabilities $P_i$ that
a next crossing occurs at the i-th iteration of the map after a separatrix 
crossing.
We note that $P_i$ does not depend on $\eps M_0(T)$ and $W_h$ for small $i$ 
because $W_h$ is significantly larger than $\eps M_0(T)$ in our consideration.
Due to the geometry of the separatrices dynamics of $H$ leads to 
dynamics of fluid particles from one cell to another. Since each cell has four 
separatrices, the probabilities for \spr crossings $P_i$, give a set of new 
probabilities $\tP_1, \tP_2, \tP_3$, and $\tP_4$ for a particle moving from
one cell to the neighbor through one of four separatrices enclosing a cell 
(see Figure 5).

	The random phase assumption is not correct in the region 
with elliptic islands or cantori. Therefore, it follows that the tails of 
the distribution of $P_i$'s for the random
model and the separatrix map can be different since there exist orbits confined in
one cell for a very long time due to the stickiness of elliptic islands 
and the effect of cantori as partial barriers for transport. But 
the $\tP_i$'s are 
almost independent of the structure of the layer since the tails of the 
distribution of $P_i$ contributes almost uniformly to $\tP_i$, so that $\tP_i$ 
is mostly
determined by the distribution of $P_i$ for small $i$ which are related only to 
the dynamics near the separatrix. Table 1 provides a comparison between values of
$P_i$ and $\tP_i$ computed numerically by the separatrix map, the direct 
integration and from the random model (20) for various values of $T$ with 
$\eps = 0.02$. Note that the dependence of $P_i$ and $\tP_i$ on $T$ 
is small and two set of values from the separatrix map and from the numerical
integrations show excellent agreements.

	The differing values of $\tP_i$'s in Table 1 implies that the random 
process over the cell is not a random walk but a {\em correlated random walk} in
which the transition probabilities depend on the entering direction and the 
rotating direction of the cell. The fluid particle is more likely to exit
through the \spr which it passes by earlier. This correlation over two 
successive steps allows
that fluid particles persist to move in the similar direction along the zig-zag
path for a few steps.
The effect of correlation conspicuously appears in the speed of the diffusion, so 
that the correlated random walk leads to modification of the diffusion constant.
The ratio of diffusion constants between the correlated and uncorrelated random 
walks is called the {\em correlation factor} denoted by $f_c$ \cite{HsKhr}. 
If the (u,v) are the coordinates of a particle after N random walk steps, the 
mean square distance of the two-dimensional correlated random walk is given by
	\bg
	L_{ms} = \left\langle u^2 + v^2 \right\rangle 
			= f_c L_s^2 N
	\ed
where $L_s$ is a length of one random walk step and the bracket represents the 
ensemble average. The random model probabilities $\tP_1, \tP_2, \tP_3, \tP_4$ 
in Table 1 yield $f_c$ $\simeq$ 1.4 with $L_s$ $\> {1\over 2}$. 

\subsection{The calculation of the diffusion constant}
In order to 
know the time dependence
of $L_{ms}$ we need to know the average period of chaotic oscillations within 
the stochastic layer, $\tau_{av}$, and the average number of random walk 
steps of particles after $J$ iterations of the separatrix map, 
$n_{rw}(J)$.  If we know these quantities, a straightforward argument gives
	\bg
	L_{ms} = f_c L_s^2 \frac{n_{rw}(J)}{J}\frac{4}{\tau_{av}}t
	\ed
and the diffusion constant $D$ 
	\bg
	D = \frac{L_{ms}}{t} = f_c L_s^2 \frac{n_{rw}(J)}{J}\frac{4}{\tau_{av}},
	\ed
where $\tau_{av}$ is obtained by averaging $T(H)$ over the separatrix layer
	\bq
	\tau_{av} & = & \frac{1}{W_h}\int_{0}^{W_h}\,\frac{2}{\pi}
						\ln{2\over \pi H} \, dH \cr
		& = & \frac{2}{\pi} \left[ 1 - 
			\ln{\frac{\eps\pi}{2(\sqrt{10}-\sqrt{6})T^2}}
			+ \ln{\sinh{\pi\over 2T}} \right]. 
	\eq
In order to perform an accurate calculation we need to exclude the elliptic islands
within the layer, so that Equation (37) yields values slightly smaller than 
accurate ones, but this difference is negligible in calculating the 
diffusion constant. 
Remarkably, $n_{rw}(J)/J$ is determined only by the ratio between the area of 
the stochastic layer and the area crossing the separatrix under one iteration 
of the separatrix map, so that we get  
	\bg
	\frac{n_{rw}(J)}{J} = \frac{2\eps M_0(T)T}{\pi S(\eps,T)},
	\ed
where $S(\eps, T)$ is the area of the stochastic layer in the phase space 
of the separatrix map. This fact is proved in Appendix. This property
implies that the global diffusion constant  is not affected by the stickiness 
of elliptic islands and the cantori in the stochastic layer except the 
area considerations of the elliptic islands in the phase space of the 
separatrix map. 

	We compute $S(\eps, T)$ by subtracting the area of 
the elliptic islands within stochastic layer from the area of the region between
two boundary circles determining the \stch layer, $2W_h T$. 
Therefore, we define a quantity by 
	\bg
	\sigma(\eps, T) = \frac{1}{\sqrt{10} - \sqrt{6}} \frac{S(\eps,T)}{W_h T}
		= \frac{S(\eps,T) T}{\eps} \sinh{\left({\pi \over 2T}\right)},
	\ed
which is proportional to the ratio between the total area of the \stch layer
without elliptic islands and $2W_hT$. When $T$ is small $\sigma(\eps, T)$ 
does not depend much on the parameters. If $T$ is not sufficiently small, then 
$\sigma (\eps,T)$ depends on $T$ in a complicated manner. Table 2 shows numerically
computed values of $\sigma (\eps, T)$ for various values of $T$, which indicates that
$\sigma(\eps, T)$ is roughly constant. This table also provides a comparison 
between the values of $n_{rw}(J)/J$ obtained by the relation  \bg
	{n_{rw}(J) \over J} = \frac{T}{\sigma (\eps,T) \pi}
	\ed
and the numerical evaluations by the direct integration, which confirms the 
validity of Equation (38).

The numerical observations in Table 2 suggests that $\sigma(\eps, T) \simeq 2.3$
, so that we get 
	\bg
	S(\eps, T) \simeq 2.3 \frac{\eps}{T} \csch{\pi \over 2T}.
	\ed
Therefore, the analytic estimates of $L_{ms}$ and $D$ are given by
	\bq
	L_{ms} & = & f_c L_s^2 \frac{T}{\sigma (\eps,T) \pi} \frac{4}{\tau_{av}} t \cr
		& \simeq & \frac{f_c}{2.3} \frac{Tt} {2\left[ 1 -
			\ln{\frac{\eps\pi}{2(\sqrt{10}-\sqrt{6})T^2}}
                        + \ln{\sinh{\pi\over 2T}} \right]}
	\eq
and
	\bg
	D = \frac{L_{ms}}{t} = \frac{f_c}{2.3} \frac{T} {2\left[ 1 -
		\ln{\frac{\eps\pi}{2(\sqrt{10}-\sqrt{6})T^2}}
			+ \ln{\sinh{\pi\over 2T}} \right]},
	\ed
with the correlation factor $f_c \sim 1.4$ from Table 1.

Figure 6 provides a comparison of $L_{ms}$ obtained by numerical simulations 
with the separatrix map and the direct integration using the 
fourth order Runge-Kutta method with the time step of 0.005. The ensemble average is done over 1900
fluid particles initially  distributed uniformly in the stochastic layer for 
the direct 
integration and over 10000 fluid particles for the separatrix map with 10000 
iterations of the separatrix map.
	The $L_{ms}$ from direct integration shows larger fluctuations due to
the smaller number of ensembles taken for the average. After same transient
behavior, the slope of the two curves converge, which demonstrates the good 
agreements between results of the \spr map and the direct numerical 
integrations.

In order to get an accurate value of the diffusion constant, fluid particles
must be distributed uniformly in the stochastic layer. But due to the effect of the 
islands and the cantori it may take very long time for fluid particles 
starting near the separatrix to fill the whole \stch layer uniformly. Numerical 
results indicate that it needs at least
$10^6$ iterations for our parameter values to achieve this uniformity, which
results in the extremely slow convergence of the diffusion constant, as seen
in Figure 7. 
Consequently, it is difficult to get the accurate diffusion 
constant by the direct integration with our computation power since 
computational efforts increase as $T$ decrease. Therefore the separatrix map
is used heavily in studying the diffusion numerically. The initial values
of diffusion constants from direct integrations and the iterations of the
\spr map are shown in the inset of Figure 7, which are good agreements and 
exhibit similar convergence behaviors.
Figure 8 shows the comparison between the analytical results from Equation (43)
and the numerical ones from iterations of the \spr map. The 
numerical results are 
obtained from the ensemble average over 10000 fluid particles distributed in 
the stochastic layer and over $10^6$ iterations of the separatrix map. 
The analytically and numerically computed diffusion constants show excellent
agreements, confirming the validity of our assumptions.

\section{Conclusion and discussions}

In the planar periodic vortical flow described by the time-dependent stream
function, fluid particles can be transported along the globally connected 
stochastic web. By considering the separatrix map  and applying the theory of 
the twist maps to the separatrix map we get analytic expressions for the 
width of the stochastic layer and the global diffusion constant, which is in 
good agreements with numerical simulations.

	In particular, when $T$ is small, there are several advantages
of using the separatrix map. Although the separatrix map yields a large error 
in the phase $t$
near the separatrix, it gives  a quite accurate estimate for the diffusion
constant because the random phase assumption is available near the separatrix.
Remarkably, the long-time correlations near the invariant circles or the cantori
do not have a direct influence on the diffusion constant except that their
areas have to be counted out. This fact was also mentioned 
by Lichtenberg and Wood \cite{Lch} in the context of the random model. The 
random phase assumption leads to
the global transport modeled by correlated random walks from one
cell to another. In fact, the difference of the diffusion constant between the
original Hamiltonian system and the separatrix map is mainly due to the errors 
in the transition probabilities of the correlated random walks, the average
period of chaotic oscillations $\tau_{av}$, and 
the areas of the phase space available for global transport and the regions 
that cross the separatrix by
one iteration of the map, which can be obtained accurately with very small
errors (from Table 2). 

For the case of small $\eps$ and $T$, the diffusion is a normal Einstein 
diffusion and no accelerator mode has been found, the existence of which 
changes the nature
of the diffusion. It should be interesting to study the limit of large $\eps$
to pursue the possibility of finding these modes by looking at the phase 
portraits and numerically computing the diffusion constants as a function 
of $\eps$ \cite{Och}.

For the area-preserving maps, two methods have  been 
widely used for studying transport. One is the Markov chain modeling of the 
mechanism of transport of phase space points suggested by Mackay {\em et al.}
\cite{Mck, MssOtt} and the other is the lobe dynamics developed by Rom-Kedar 
{\em et al.} \cite{RkdLnrdWg, RkdWg}. 
When $T$ is small, the separatrix 
map and the Melnikov theory reveal that there are many relatively large elliptic
islands in the stochastic layer. Since these islands act as partial barriers
for transport and divide a phase space into many partitions it is a good 
situation for a Markov 
chain modeling with which one can compute the statistical quantities
such as the probability distribution of the trapping times in a cell. 
With the lobe dynamics, it is 
more difficult to study the detailed dynamics in one cell such as the statistics
of the trapping time and long time correlations near islands since it needs a
very large number of iterations for the turnstile lobe to fill sufficiently 
the whole stochastic layer. But the lobe dynamics allows one to study the dynamics 
near the \spr without the random phase assumption. Though we have focused on
the global transport, it should be also interesting to study the local transport 
within the \stch layer by using the \spr map in conjunction with the Markov
chain models and the lobe dynamics.
\vspace*{1cm}

{\Large \bf Acknowledgements}  \\
We would like to thank Prof. Yup Kim for helpful remarks and Prof. H. Mori 
for valuable comments on the paper. This
work was supported in part by the Basic Science Research Institute under the 
contract No. N91126 and in part by KOSEF through the Global Analysis Research 
Center.

\vspace*{1cm}

\renewcommand{\theequation}{A\arabic{equation}}
\setcounter{equation}{0}

{\Large \bf Appendix}  \\
Let $S$ be an area preserving map on the cylinder $T^1 \times \FR = (\FR / \FZ)
\times \FR $ and $U$ be an open invariant set with finite area. For a rotational
circle $C$ contained in $U$ we define the subset $A$ of $U$ by the set of points
which cross $C$ by one iteration of $S$, that is, $x$ and $Sx$ are located on 
different sides of $C$. The average number of crossing $C$ after $J$ iterations
of $S$, denoted by $n_{av}(J)$, is given by 
	\bg
	n_{av}(J) = \frac{1}{\mid U \mid} \int_{U} n(x; J) \, dx
	\ed
where $n(x; J)$ is the frequency that a point $x$ crosses $C$ while $S$ is iterated 
$J$ times and $\mid U\mid$ is the area of U. 

From the definition of $A$ we have a recurrent relation on $n(x; J)$  
	\bq
	n(x; J) & = & n(x; J-1) + 1 \mbox{\hspace{2cm} if} \>\>\> x \in A \cr
		& = & n(x; J-1) \mbox{\hspace{3cm} if} \>\>\> x\notin A
	\eq
and, in particular, when $J=1$
	\bq
	n(x; 1) & = & 1 \mbox{\hspace{2cm} if} \>\>\> x \in A \cr
		& = & 0 \mbox{\hspace{2cm} if} \>\>\>  x \notin A. 
	\eq
Thus $n_{av}(J)$ has a following recurrent relation
	\bq
	n_{av}(J) & = & \frac{1}{\mid U \mid} \int_{U \backslash A} n(x; J-1) 
		\,dx + \frac{1}{\mid U \mid} \int_{A}{n(x; J-1) + 1} \, dx \cr
		& = & n_{av}(J-1) + \frac{\mid A \mid}{\mid U \mid} 
	\eq
with the initial condition 
	\bg
	n_{av} (1) = \frac{\mid A \mid}{\mid U \mid}.
	\ed
Consequently, we have  
	\bg
	n_{av}(J) = J \frac{\mid A \mid}{\mid U \mid}.
	\ed
For our system, we have 
	\bg
	\mid U \mid = S(\eps, T) \mbox{\hspace{1cm} and \hspace{1cm}} 
	\mid A \mid = \frac{2T}{\pi}\eps M_0(T) ,
	\ed
which yields 
	\bg
	\frac{n_{rw}(J)}{J} = \frac{2 \eps M_0(T)T}{\pi S(\eps,T)}. 
	\ed

\newpage
{\Large \bf Figure captions } \\

\begin{center}
        \begin{minipage}[h]{15cm}
           {Figure 1: The phase space of the unperturbed system in 
		Equation (5) with $\eps = 0$. The system is doubly periodic in 
		$x$ and $y$ and the separatrices are globally connected.} \\
	\vspace*{1cm}
        \end{minipage}

        \begin{minipage}[h]{15cm}
           {Figure 2: A sketch for the construction of the separatrix map. The
		orbit $q_s(t)$ is the heteroclinic orbit of the unperturbed
		system and $q(t)$ a perturbed orbit. The sections $\Sigma^H$
		and $\Sigma^t$ are the Poincar\'{e} sections for the 
		coordinates $(H, t)$.  \\}
	\vspace*{1cm}
        \end{minipage}

        \begin{minipage}[h]{15cm}
                {Figure 3: The phase portrait of the time-$T$ Poincar\'{e}
                        map near a saddle point is shown in (a) and the
                        phase portrait of the composite \spr map $S$ in 
			(b) with $\eps = 0.02$, and $T = 0.085$.
			These two phase portraits, though in different 
			coordinate systems, show clearly the 
			correspondence between the two island structures.  \\}
	\vspace*{1cm}
        \end{minipage}

        \begin{minipage}[h]{15cm}
                {Figure 4: The width of the stochastic layer with 
			$\eps = 0.02$. The analytical results from Equation 
			(31) are shown in a solid line and the numerical 
			results from the direct integration and the 
			iterations of the \spr map in squares and crosses, 
			respectively.  \\} 
	\vspace*{1cm}
        \end{minipage}

        \begin{minipage}[h]{15cm}
                {Figure 5: The correlated random walk model for the global
		transport describing the cell to cell dynamics. The dashed 
		direction represents the entering direction of a fluid 
		particle into a cell and $\tP_i$ are the 
		transition probabilities to the neighboring cells.  \\}
	\vspace*{1cm}
        \end{minipage}

        \begin{minipage}[h]{15cm}
                {Figure 6: The plot of the number of iterations, $n$, 
			versus $L_{ms}$ from 
			the direct integrations and the \spr map with 
			$\eps = 0.02$ and $T = 0.125$. The curve from the
			direct integrations (below) shows more fluctuations 
			since the size of the ensemble is five times 
			smaller than one from the \spr map.  \\}
	\vspace*{1cm}
        \end{minipage}

	\begin{minipage}[h]{15cm}
		{Figure 7: The plot of the number of iterations n versus the
			diffusion constant $D$ computed from the \spr map 
			when $\eps = 0.02$ and $T = 0.1$, which shows the 
			extremely slow convergence. The inset shows the 
			initial comparison between the diffusion constants
			from the numerical integration and the \spr map. \\}
	\vspace*{1cm}
	\end{minipage}

        \begin{minipage}[h]{15cm}
                {Figure 8: The diffusion constant computed from Equation (43)
		is shown in a solid line and the one from iterations of the
		\spr map in diamonds, which are in excellent agreements.  \\} 
        \end{minipage}
\end{center}

\newpage

\begin{table}[t]
  \begin{center}
  	\caption{The probabilities of the separatrix crossing at the $i$th 
		iteration $P_i$ and the probabilities of the fluid particle 
		jumping from one cell to another, $\tP_i$, $i$ = 1, 2, 3, 4 
		are computed by numerical simulations with the separatrix map 
		and direct integrations. They are compared with values given by
		 random model in (20), which shows good agreements.}
	\begin{tabular}{|c|c|c|c|c|c|c|c|c|c|} \hline
		$\eps$ = 0.02 & T & $P_1$ & $P_2$ & $P_3$ & $P_4$ & 
				$\tP_1$ & $\tP_2$ & $\tP_3$ & $\tP_4$ \\ \hline
		The & 0.085 & 0.362 & 0.113 & 0.073 &0.048 &
				0.482 & 0.218 & 0.168 & 0.132 \\
		separatrix & 0.1 & 0.366 & 0.112 & 0.073 & 0.050 &
				0.489 & 0.214 & 0.165 & 0.132 \\
		map	&  0.16 & 0.368 & 0.109 & 0.090 & 0.043 & 
				0.479 & 0.209 & 0.183 & 0.129 \\ \hline
		The & 0.085 & 0.362 & 0.113 & 0.073 & 0.048 &
				0.481 & 0.215 & 0.165 & 0.130 \\
		numerical & 0.1 & 0.364 & 0.111 & 0.073 & 0.050 &
				  0.487 & 0.211 & 0.162 & 0.131 \\
		integration & 0.16 & 0.366 & 0.109 & 0.088 & 0.043 &
				0.478 & 0.209 & 0.181 & 0.130 \\ \hline
		Random model & & 0.363 & 0.112 & 0.075 & 0.048 &
				0.480 & 0.215 & 0.170 & 0.135 \\ \hline
	\end{tabular}
  \end{center}
\end{table}

	\begin{table}[b]
	  \begin{center}
		\caption{The area $S(\eps, T)$ is computed numerically 
		with $\eps = 0.02$, from which 
		$n_{rw}(J)/J$ is computed by using Equation (38) and compared 
		with the results of numerical simulations. \\}  
		\begin{tabular}{|c|c|c|c|} \hline
		  T & $\sigma(\eps,T)$ & $\frac{1}{J}n_{rw}(J)$ from Equation 
		  (25) & numerical values of $\frac{1}{J}n_{rw}(J)$ \\ \hline
			0.075 & 2.270 & 0.01052 & 0.01075 \\
			0.1   & 2.238 & 0.01422 & 0.01475 \\
			0.13  & 2.433 & 0.01701 & 0.01778 \\
			0.16  & 2.265 & 0.02249 & 0.02299 \\ 
			0.2   & 2.363 & 0.02694 & 0.02786 \\ \hline
		\end{tabular}
	  \end{center}
	\end{table}

\end{document}